\documentclass[twoside,slac_one]{revtex4}
\usepackage{graphicx}
\usepackage{fancyhdr}
\usepackage{amsmath} 
\usepackage{bm}
\usepackage{amsxtra}
\usepackage{amssymb}
\usepackage{amsthm}
\usepackage{latexsym}
\usepackage{lscape}

\newcommand{\be}{\begin{equation}}
\newcommand{\ee}{\end{equation}}
\newcommand{\ba}{\begin{eqnarray}}
\newcommand{\ea}{\end{eqnarray}}
\newcommand{\fnl}{f_{\rm NL}}
\newcommand{\gnl}{g_{\rm NL}}
\newcommand{\x}{{\bf x}}
\newcommand\lsim{\mathrel{\rlap{\lower4pt\hbox{\hskip1pt$\sim$}}
        \raise1pt\hbox{$<$}}}
\newcommand\gsim{\mathrel{\rlap{\lower4pt\hbox{\hskip1pt$\sim$}}
        \raise1pt\hbox{$>$}}}
\begin{document}

\title{Testing Inflation with Dark Matter Halos}

%

\author{Marilena LoVerde}
\affiliation{Institute for Advanced Study, Princeton, NJ USA}

\author{Simone Ferraro \& Kendrick M.~Smith}
\affiliation{Department of Astrophysical Sciences, Princeton University, Princeton NJ USA}

\begin{abstract}
Cosmic inflation provides a mechanism for generating the early density perturbations that seeded the large-scale structures we see today. Primordial non-Gaussianity is among the most promising of few observational tests of physics at this epoch. At present non-Gaussianity is best constrained by the cosmic microwave background, but in the near term large-scale structure data may be competitive so long as the effects of primordial non-Gaussianity can be modeled through the non-linear process of structure formation.  We discuss recent work modeling effects of a few types of primordial non-Gaussianity on the large-scale halo clustering and the halo mass function. More specifically, we compare analytic and N-body results for two variants of the curvaton model of inflation: (i) a ``$\tau_{NL}$" scenario in which the curvaton and inflaton contribute equally to the primordial curvature perturbation and (ii) a ``$g_{NL}$'' model where the usual quadratic $f_{NL}$ term in the potential cancels, but a large cubic term remains. 
\end{abstract}

\maketitle

\thispagestyle{fancy}


\section{Introduction}
From the anisotropies in the cosmic microwave background (CMB) to the large-scale distribution
 of dark matter halos hosting galaxies, the universe appears rich with structure. A central goal of observational 
 cosmology is to understand the period of inflation that is believed to have generated this structure
 \cite{Guth:1982ec,Hawking:1982cz,Starobinsky:1982ee,Bardeen:1983qw}. Current CMB data 
 confirms inflationary predictions for a spatially flat universe with primordial curvature perturbations drawn
 from a nearly scale-invariant power spectrum \cite{Komatsu:2010fb}. Nevertheless, distinguishing between microphysical 
 models remains a challenge. 
 
 Evidence for non-Gaussianity in the primordial perturbations could rule out large classes of inflationary 
 models and shed light on the mechanism that generated the initial structure (see e.g. \cite{Bartolo:2004if, Chen:2010xka}). At present
the most stringent bounds on primordial non-Gaussianity come from CMB experiments \cite{Komatsu:2010fb,Fergusson:2010gn,Smidt:2010sv}, 
but data from galaxy surveys is increasingly competitive (see, e.g.  \cite{Slosar:2008hx,Carbone:2010sb,Shandera:2010ei}). In these proceedings 
we give an overview of analytic and N-body results for the abundance and clustering of dark matter halos with local non-Gaussian initial conditions
 described by the parameters $\fnl$, $\gnl$ and $\tau_{NL}$.  In \S \ref{sec:ICs} we introduce three examples of non-Gaussian initial conditions. 
 In \S \ref{sec:analytic} we review analytic models of the impact of local non-Gaussian initial conditions on the abundance and clustering 
 of dark matter halos. Comparison of the analytic models and results from N-body simulations is given \S \ref{sec:nbody}. 
In \S \ref{sec:summary} we summarize our results. The reader is referred to the original references 
\cite{Smith:2010gx,LoVerde:2011iz,Smith:2011ub,Smith:2011if}
 for a more detailed discussion.

\section{Examples of local non-Gaussian statistics: $\fnl$, $\gnl$, $\tau_{NL}$}
\label{sec:ICs}
If the initial curvature\footnote{Following standard notation in studies of non-Gaussianity, we define $\Phi = \frac{3}{5} \zeta$,
where $\zeta$ is the gauge invariant primordial curvature.} perturbations are homogeneous, 
isotropic and Gaussian their statistics are entirely characterized by 
the two point correlation function, or power spectrum
\be
\langle \Phi({\bf k})\Phi({\bf k'})\rangle =(2\pi)^3 \delta_{{\rm Dirac}}({\bf k}+{\bf k'})P_{\Phi\Phi}(k)\,.
\ee
If $\Phi$ has any non-zero odd $N$-point function or even $N$-point function that just isn't specified by the two-point function
(i.e. $\langle \Phi(\x_1)\Phi(\x_2)\dots \Phi(\x_N)\rangle\neq\langle\Phi(\x_1)\Phi(\x_2)\rangle\langle\Phi(\x_3)\Phi(\x_4)\rangle\dots\langle\Phi(\x_{N-1})\Phi(\x_N)\rangle +{\rm permutations}$)
then $\Phi$ is non-Gausssian. 

While there are an abundance of inflationary scenarios producing a variety of non-Gaussian 
initial conditions (there are some organizing principles; see for example~\cite{Babich:2004gb}) here we focus on 
initial curvature that can be written as a non-linear mapping of a Gaussian random field that is local in real space. These types of 
initial conditions can arise in the curvaton model \cite{Linde:1996gt,Lyth:2001nq}. For a general review of inflationary scenarios
giving local non-Gaussian initial conditions see \cite{Byrnes:2010em}.  
 
\subsection{$\fnl$} 
\label{ssec:fnl}
Consider defining the initial curvature as a Gaussian random field $\Phi_G$, plus a small ($\mathcal{O}(\fnl \sqrt{\langle \Phi_G^2\rangle})$) 
fractional perturbation \cite{Komatsu:2001rj}
 \be
\label{eq:fnldef}
\Phi_{NG}(\x)=\Phi_G(\x)+\fnl\left(\Phi_G^2(\x)-\langle \Phi_G^2\rangle\right)\,.
\ee
The new field $\Phi_{NG}(\x)$ obeys non-Gaussian statistics. In particular it has a skewness $\langle \Phi_{NG}^3\rangle \sim 6\fnl \langle \Phi_G^2\rangle^2$, and kurtosis
$\langle \Phi_{NG}^4\rangle_c\equiv \langle \Phi_{NG}^4\rangle-3\langle \Phi_{NG}^2\rangle^2\sim 48 \fnl^2\langle \Phi_G^2\rangle^3$. The CMB constraints on this
form of primordial non-Gaussianity are $-10<\fnl<74$ at $95\%$ confidence \cite{Komatsu:2010fb}.

Another non-Gaussian 
feature is the coupling between different physical scales. We can gain some insight into the mode coupling by splitting the Gaussian field into ``short" and ``long"
 wavelength pieces, $\Phi_G=\Phi_{G,s}+\Phi_{G,l}$.  The short wavelength fluctuations of the non-Gaussian field depend on both short and long wavelength modes of the Gaussian field, 
\be
\Phi_{NG,s}=\Phi_{G,s}+\fnl\left(\Phi_{G,s}^2-\langle \Phi_{G,s}^2\rangle\right)+2\fnl \Phi_{G,s}\Phi_{G,l}\,.
\ee
In particular, an observer on top of a long wavelength mode $\Phi_{G,l}$ will see small-scale statistics that depend on the value of $\Phi_{G,l}$
\be
\label{eq:fnlpbsplit}
\left.\langle \Phi_{NG,s}^2\rangle\right|_{l} \sim \langle \Phi_{NG,s}^2\rangle\left(1+4\fnl \Phi_{G,l}+4\fnl^2\Phi_{G,l}^2\right)
\ee
\be
\left.\langle \Phi_{NG,s}^3\rangle\right|_{l} \sim 6\fnl \langle \Phi_{G,s}^2 \rangle^2 \left(1+4\fnl \Phi_{G,l}\right)  \label{eq:fnl_local_skewness}
\ee
where we have only kept terms up to $\mathcal{O}(\fnl^2)$. This coupling between short and long wavelength scales will be important to understand the impact of non-Gaussian initial conditions 
on the clustering of dark matter halos.

\subsection{$\gnl$}
\label{ssec:gnl}
A variant on Eq.~(\ref{eq:fnldef}) is to consider a local mapping where the quadratic term vanishes, and the cubic term is important \cite{Okamoto:2002ik, Enqvist:2005pg}
\be
\label{eq:gnldef}
\Phi_{NG}(\x)=\Phi_G(\x)+\gnl\left(\Phi_G^3(\x)-3\Phi_G(\x)\langle \Phi_G^2\rangle\right)\,.
\ee
With this mapping the skewness of $\Phi_{NG}$ vanishes and the kurtosis is $\langle\Phi_{NG}^4\rangle_c \sim 24 \gnl \langle \Phi_{G}^2\rangle^3$. The CMB limits
$-12.34 < \gnl/10^5 <15.58$ at $95\%$ confidence \cite{Fergusson:2010gn}.

In the $\gnl$ model, the coupling of short and long scales in Eq.~(\ref{eq:gnldef}) gives a small-scale variance that depends on $\Phi_{G,l}^2$, 
\be
\left.\langle \Phi_{NG,s}^2\rangle\right|_{l}\sim \langle \Phi_{NG,s}^2\rangle\left(1+6 \gnl\Phi_{G,l}^2\right)
\ee
and a local skewness the varies linearly with $\Phi_{G,l}$
\be
\left.\langle \Phi_{NG,s}^3\rangle \right|_{l}\sim 18\gnl \langle \Phi_{G,s}^2\rangle ^2\Phi_{G,l}\,.   \label{eq:gnl_local_skewness}
\ee
Comparing the leading terms in Eqs.~(\ref{eq:fnl_local_skewness}) and~(\ref{eq:gnl_local_skewness}), we see that 
 to an observer sitting on a long wavelength mode $\Phi_l$, it appears that they live in a cosmology with $\fnl^{eff}(\x)=3 \gnl \Phi_{G,l}(\x)$!
\subsection{$\tau_{NL}$}
\label{ssec:tnl}

Another variation is to consider initial curvature which is the sum of two fields
$\phi_G$ and $\sigma_{G}$ which fluctuate independently ($\langle \phi_G\sigma_G\rangle=0$)
but have proportional power spectra, $P_{\phi\phi}/P_{\sigma\sigma}\equiv \xi^2 = {\rm constant}$. Non-Gaussianity is generated by adding a 
term quadratic in $\sigma$, 
\be
\label{eq:tnldef}
\Phi_{NG}(\x)=\phi_G(\x)+\sigma_G(\x)+\fnl(1+\xi^2)^2\left(\sigma_G^2(\x)-\langle \sigma_G^2\rangle\right)\,.
\ee
The skewness in this model is given by $\langle \Phi_{NG}^3\rangle \sim 6\fnl\langle \Phi_{NG}^2\rangle^2$, and the kurtosis is 
$\langle \Phi_{NG}^4\rangle_c\sim 48 \fnl^2(1+\xi^2)\langle \Phi_{NG}^2\rangle ^3\equiv 48 \tau_{NL}\langle \Phi_{NG}^2\rangle ^3/(\frac{6}{5})^2$ where
the factor of $6/5$ is conventional.\footnote{Apparently because $\fnl$ is typically defined through Eq.~(\ref{eq:fnldef}) giving $\fnl\sim \frac{1}{6}\langle \Phi^3\rangle/\langle\Phi^2\rangle^2$, where 
$\Phi=\frac{3}{5}\zeta$ with $\zeta$ the primordial curvature, but $\tau_{NL}$ is typically defined in terms of $\zeta$ as $\tau_{NL}\sim \frac{1}{12}\langle \zeta^4\rangle_c/\langle \zeta^2\rangle^2$ \cite{Boubekeur:2005fj}.} Current CMB bounds on this parameter are $-6000 <\tau_{NL} <33,000$ at $95\%$ confidence \cite{Smidt:2010sv}. 

In this $\tau_{NL}$ model, there is a coupling between small-scale fluctuations in $\Phi_{NG}$ and the long-wavelength fluctuations in $\sigma_G$
\be
\label{eq:tpbsplit}
\left.\langle \Phi_{NG,s}^2\rangle\right|_{l} \sim \langle \Phi_{NG,s}^2\rangle\left(1+4\fnl(1+\xi^2) \sigma_{G,l}+4\fnl^2(1+\xi^2)^3\sigma_{G,l}^2\right)\,.
\ee
 In this two-field example $\tau_{NL}\ge (\frac{6}{5}\fnl)^2$. One may wonder whether this inequality is fundamentally related to the physics of inflation. 
 \cite{Suyama:2007bg} showed the inequality was true at tree-level using the $\delta N$ formalism. In fact, this inequality can be interpreted as a 
 positivity constraint that must be satisfied regardless of the mechanism that generated the perturbations.  A formal proof is given in \cite{Smith:2011if}, 
 but the argument can be understood as follows. 
From Eqs.~(\ref{eq:fnlpbsplit}) \& (\ref{eq:tpbsplit}) we can see that $\fnl$ is a measure of the large-scale
correlation between the potential $\Phi$ and the locally measured small-scale power 
($\fnl \sim \langle \Phi_l \Phi_s^2\rangle/\langle \Phi_l^2\rangle\langle \Phi_s^2\rangle$).
On the other hand, $\tau_{NL}$ is a measure of the large-scale variance in the small-scale
power ($\tau_{NL}\sim \langle \Phi_l^2 \Phi_s^2\rangle_c/\langle \Phi_l^2\rangle^2\langle \Phi_s^2\rangle$).
The inequality $\tau_{NL} \ge (\frac{6}{5}\fnl)^2$ then arises as the condition that the correlation
coefficient between the small-scale power and $\Phi$ must be between -1 and 1.

\section{Impact of non-Gaussian initial conditions on large-scale structure: analytic predictions}
\label{sec:analytic}
In the previous section we discussed non-Gaussianity in the initial curvature perturbation.  To understand the impact of primordial non-Gaussianity
on the abundance and clustering of dark matter halos, we need a model that relates halos to perturbations in the initial curvature.
 In linear theory, the matter density perturbation $\delta$ is simply related to the initial curvature,
\be
\delta({\bf k},z)=\frac{2 k^2T(k)D(z)}{3 \Omega_m H_0^2}\Phi({\bf k})\equiv \alpha(k,z) \Phi({\bf k})
\ee
where $T(k)$ is the transfer function and $D(z)$ is the linear growth function. A simple prescription for halo formation is to smooth 
the linear density field on scale $M$
\be
\delta_M(\x)\equiv\frac{1}{V}\int_{V\sim M/\rho_m}\!\! d^3\x' \delta(\x-\x') 
\ee
and to model halos of mass $>M$ as regions of the smoothed initial density field with $\delta_M(\x)>\delta_c$, where $\delta_c$ is the spherical collapse threshold.
The probability distribution function (PDF) for fluctuations $\delta_M$ can be written in terms of the cumulants for the smoothed density fluctuation $\delta_M$
\be
\sigma_M^2\equiv\langle \delta_M^2\rangle\,, \quad \kappa_3(M)\equiv\frac{\langle \delta_M^3\rangle}{\sigma_M^3}\,,\quad \kappa_4(M)\equiv\frac{\langle \delta_M^4\rangle_c}{\sigma_M^4}\,,\quad\dots
\ee                                                                                    
where $\kappa_3\propto \fnl$ and $\kappa_4$ contains terms $\propto \gnl$ and $\propto \tau_{NL}$. If $\delta_M$ is non-Gaussian, an infinite number 
of cumulants are needed to completely specify the PDF. However, one can approximate the PDF with a finite set of cumulants and for our purposes
$\sigma_M^2$, $\kappa_3$ and $\kappa_4$ are sufficient.

In the next sections we use these elements to model the abundance and clustering of dark matter halos. 
In the plots throughout this paper we use the WMAP5+BAO+SN fiducial cosmology \cite{Dunkley:2008ie}:
baryon density $\Omega_bh^2 = 0.0226$, cold dark matter (CDM) density $\Omega_ch^2 = 0.114$, Hubble parameter $h=0.70$,
spectral index $n_s=0.961$, optical depth $\tau = 0.080$, and power-law initial curvature power spectrum 
$k^3 P_\zeta(k) / 2\pi^2 = \Delta_\zeta^2 (k/k_{\rm piv})^{n_s-1}$ where $\Delta_\zeta^2 = 2.42 \times 10^{-9}$
and $k_{\rm piv} = 0.002$ Mpc$^{-1}$.

\subsection{Halo abundance}
\label{ssec:massfcn}
Dark matter halos form from rare positive fluctuations in the matter density. Press \& Schechter \cite{Press:1973iz} gave a simple
 analytic model for the abundance of dark matter halos with mass $M$ in terms of the PDF 
 for the linear matter density fluctuations smoothed on scale $R=(3M/4\pi\rho_m)^{1/3}$, 
 \be
 \label{eq:dndm}
n(M)=-2 \frac{\rho_m}{M}\frac{\partial }{\partial M} \mathcal{P}(\delta_M>\delta_c,M)
 \ee
where $\delta_c\approx 1.68$ is the spherical collapse threshold and $\mathcal{P}(\delta_M>\delta_c,M)$ is the probability for $\delta_M>\delta_c$. 
Using 
\be
\mathcal{P}_{{\rm Gaussian}}(\delta_M>\delta_c,M)=\int_{\delta_c}^{\infty}d\delta_M \frac{1}{\sqrt{2\pi \sigma_M^2}}e^{-\frac{1}{2}\delta_M^2/\sigma_M^2}
\ee
in Eq.~(\ref{eq:dndm}) gives the Press-Schechter mass function \cite{Press:1973iz}, which is known to disagree at the $\sim 50\%$ level with the mass function 
measured from N-body simulations \cite{Jenkins:2000bv}. Nevertheless, the expression in Eq.~(\ref{eq:dndm}) has long been used to predict the relative
abundance of halos in a non-Gaussian cosmology to a Gaussian one \cite{Lucchin:1987yv}. Typically, the full non-Gaussian PDF 
is not known, but approximate expressions for $n(M)$ are obtained by truncating an asymptotic expansion (e.g. \cite{Matarrese:2000iz}) or Edgeworth
series (e.g. \cite{LoVerde:2007ri} -- hereafter the ``Edgeworth mass function") for the PDF and using Eq.~(\ref{eq:dndm}). These expressions agree well with N-body simulations with $\fnl$-type initial
 conditions provided the ``modified" collapse threshold $\delta_c' \approx 1.42$ is used (e.g. \cite{Pillepich:2008ka}). 

In \cite{LoVerde:2011iz}, truncating the series for $\ln \mathcal{P}(\delta_M>\delta_c,M)$ rather than $\mathcal{P}(\delta_M>\delta_c)$ was proposed, where the Edgeworth expression is used for the PDF. In the limit
of small non-Gaussian corrections, the mass function obtained this way (which we call the ``log-Edgeworth" mass function) agrees with the Edgeworth mass function, 
but the in high-mass limit where non-Gaussian corrections are important, the ``log-Edgeworth" mass function is better behaved (see Figs.~(\ref{fig:fnlmassfcn}) \& (\ref{fig:gnlmassfcn})).

\subsection{Scale-dependent halo bias from $\fnl$ and $\gnl$-type initial conditions}
\label{ssec:scalebias}
In the previous section we discussed modeling dark matter halos as regions where the linear density field $\delta_M$ exceeds $\delta_c$. 
If a fluctuation $\delta_M$ is sitting on top of a longer wavelength fluctuation in the density $\delta_l$, then the local collapse threshold is adjusted to 
$\delta_c-\delta_l$, thus the halo abundance fluctuates with the density as
\be
\label{eq:deltanG}
\frac{\delta n(\x)}{n}=1+\left.\frac{\partial \ln n(M, \delta_c-\delta_l)}{\partial \delta_l}\right|_{\delta_l=0}\delta_l(\x)\,,
\ee
where the $1$ accounts for the fact that the Eulerian halo number density is increased by a factor of $1+\delta$ with respect to the Lagrangian one given by $n(M)$. 

As we've seen in \S \ref{sec:ICs} non-Gaussian initial conditions can cause small scale statistics like the variance and skewness to be modulated 
by the long-wavelength potential $\Phi_l$. For the $\fnl$ and $\gnl$ initial conditions the small scale variance and skewness are modulated by $\Phi_l$. 
Accounting for this modifies Eq.~(\ref{eq:deltanG}) to
\be
\frac{\delta n}{n}(\x)=1+\frac{\partial \ln  n}{\partial \delta_l}\delta_l(\x)+4\fnl \frac{\partial \ln n}{\partial\ln \sigma_M^2} \Phi_l(\x)+3\gnl \frac{\partial \ln n}{\partial \fnl}\Phi_l(\x)\,.
\ee
In Fourier space, the density field is related to the early-time gravitational potential through $\alpha(k,z)$ which allows us to write
\begin{eqnarray}
\label{eq:bkfg}
\frac{\delta n}{n}({\bf k})&=&\left(b+\frac{4\fnl}{\alpha(k,z)} \frac{\partial \ln n}{\partial \ln \sigma_M^2}+\frac{3\gnl}{\alpha(k,z)}\frac{\partial\ln n}{\partial \fnl}\right)\delta_l({\bf k})\\
&\equiv& b_{\fnl,\gnl}(k)\delta_l({\bf k})
\end{eqnarray}
where $b\equiv1+\partial \ln n/\partial \delta_l$. We'll refer Eq.~(\ref{eq:bkfg}) to as the ``peak-background-split" (PBS) prediction for the scale-dependent non-Gaussian bias. 

For a mass function dependent only on the combination $\delta_c/\sigma_M$, rather than $\delta_c$ and $\sigma_M$ separately, 
we can write the $\sigma_M^2$ derivative in terms of the constant bias $b$ 
\be
\label{eq:bkbfg}
\frac{\delta n}{n}({\bf k})=\left(b+\frac{2\fnl\delta_c(b-1)}{\alpha(k,z)}+\frac{3\gnl}{\alpha(k,z)}\frac{\partial\ln n}{\partial \fnl}\right)\delta_l({\bf k})\,.
\ee
The $\gnl$ bias coefficient can be rewritten in terms of $b$, but the expression is more complicated \cite{Smith:2011ub}. The $\fnl$ dependent 
bias above was first written down in \cite{Dalal:2007cu}, but the derivation given here follows that of \cite{Slosar:2008hx}. The 
expression for $\gnl$ bias term is from \cite{Smith:2011ub}, but see also \cite{Desjacques:2009jb,Desjacques:2011mq}. 

From Eq.~(\ref{eq:bkbfg}) the large-scale halo-matter and halo-halo power spectra are given by
\be
\label{eq:PmhPhhfg}
P_{hm}(k)=b_{\fnl,\gnl}(k)P_{mm}(k) \quad {\rm and } \quad P_{hh}(k)=b_{\fnl,\gnl}^2(k)P_{mm}(k)\,.
\ee
The analytic form for the $\fnl$-dependence of $P_{hm}$ and $P_{hh}$ given by Eqs.~(\ref{eq:bkbfg}) \& (\ref{eq:PmhPhhfg}) has been shown to be in excellent agreement with simulations (see e.g. \cite{Dalal:2007cu,Grossi:2009an,Pillepich:2008ka}). Using the scale dependent halo bias, \cite{Slosar:2008hx} constrain $-29 < \fnl< 70$ at $95\%$ confidence. 
Using the fact that the scale-dependent bias from $\gnl$ has the same $k$-dependence ($\propto 1/\alpha(k,z)$), \cite{Desjacques:2009jb} applied the $\fnl$ constraints from \cite{Slosar:2008hx} to limit $-3.5 <\gnl/10^5<8.2$. 

\subsection{Stochasticity between halos and dark matter from $\tau_{NL}\neq (\frac{6}{5}\fnl)^2$ initial conditions}
\label{ssec:stochastic}
For the two-field initial conditions given in \S \ref{ssec:tnl},
the linear density field is determined by the sum of the potentials $\sigma+\phi$, through 
\be
\delta({\bf k},z)=\alpha(k,z)\left(\phi({\bf k})+\sigma({\bf k})\right)\,, 
\ee
but the small scale power is modulated by $4\fnl(1+\xi^2)\sigma_{l,G}$.  
So the number of halos fluctuates as 
\be
\frac{\delta n}{n}({\bf k})=b\delta_l({\bf k})+2\fnl(1+\xi^2)\delta_c(b-1)\sigma_l({\bf k})\,.
\ee
The halo-matter cross-power spectrum is unchanged from the $\fnl$ case, 
\ba
P_{hm}(k,z)&=&\left(b+\frac{2\fnl \delta_c(b-1)}{\alpha(k,z)}\right)P_{mm}(k,z)\\
&=&b_{\fnl}(k)P_{mm}(k,z)
\ea
but the halo-halo power spectrum is now
\be
\label{eq:Phhtnl}
P_{hh}(k,z)=b_{\fnl}^2P_{mm}(k,z)+4(\frac{25}{36}\tau_{NL}-\fnl^2)\delta_c^2(b-1)^2P_{\Phi\Phi}(k)\,.
\ee
The second term above represents stochasticity of the halo field with respect to the dark matter field \cite{Tseliakhovich:2010kf}. 

\section{Impact of non-Gaussian initial conditions on large-scale structure: comparison with simulations}
\label{sec:nbody}

To study the halo mass function and clustering, we
performed collisionless $N$-body simulations using the GADGET-2 TreePM code \cite{Springel:2005mi}.
Simulations were done using periodic box size $R_{\rm box} = 1600$ $h^{-1}$~Mpc, particle count $N_p = 1024^3$, and force softening
length $R_s = 0.05 (R_{\rm box}/N_p^{1/3})$.
With these parameters and the fiducial cosmology from \S\ref{sec:analytic}, the particle mass 
is $m_p = 2.92 \times 10^{11}$ $h^{-1}$~$M_\odot$. Non-Gaussian initial conditions were implemented by 
generating Gaussian fields and applying the maps in Eq.~(\ref{eq:fnldef}), Eq~(\ref{eq:gnldef}), or Eq.~(\ref{eq:tnldef}). 
The non-Gaussian fields were linearly evolved to the initial simulation redshift, $z=100$, using the transfer functions 
from CAMB \cite{Lewis:1999bs}. Halos were identified using the Friends of Friends algorithm \cite{Frenk:1988zz} with linking length $L_{FoF}=0.2 R_{box}N_p^{-1/3}$ 
and halo positions are identified using the mean of the particle positions. For further details see \cite{Smith:2010gx,LoVerde:2011iz, Smith:2011ub}. 
 
 \begin{figure}[t]
\centering
\begin{tabular}{cc}
\includegraphics[width=80mm]{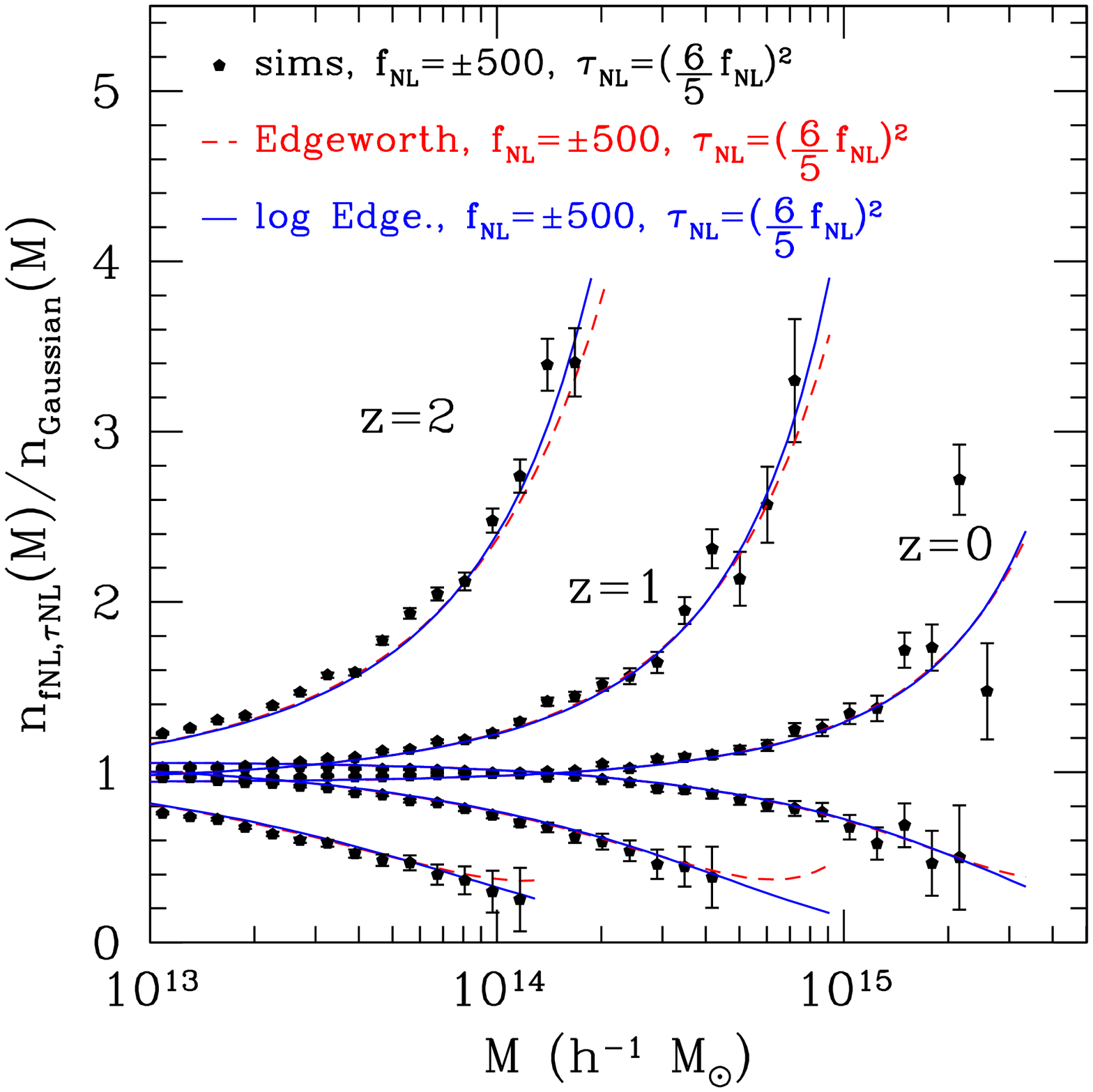}&\includegraphics[width=80mm]{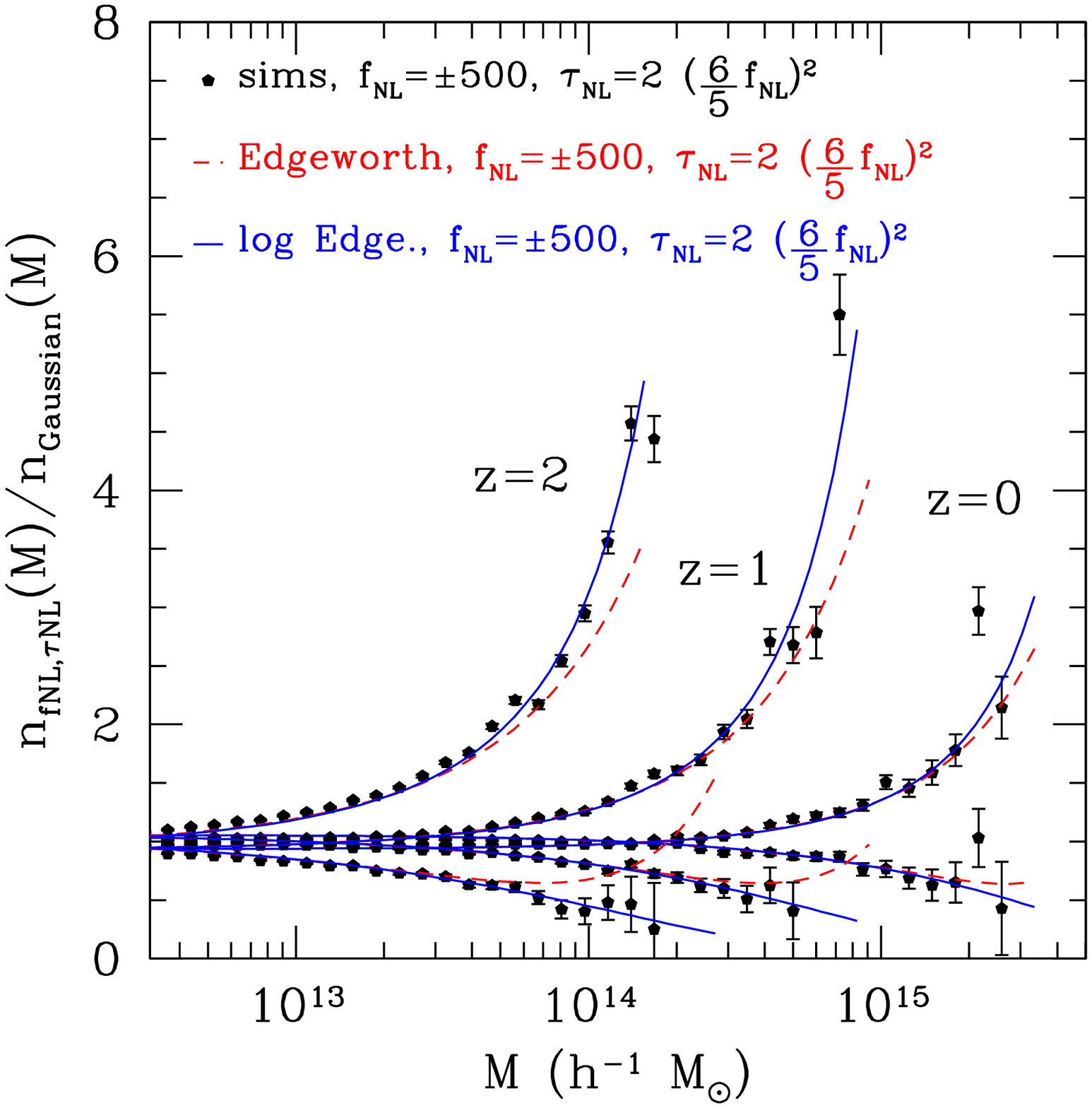}\\
\mbox{\bf(a)} & \mbox{\bf (b)}
\end{tabular}
\caption{The non-Gaussian correction to the halo mass function. Plotted is $n_{NG}(M,z)/n_G(M,z)$ for (a) $\fnl =\pm500$, $\tau_{NL}=(\frac{6}{5}500)^2$,
 (b) $\fnl =\pm 500$, $\tau_{NL}=2(\frac{6}{5}500)^2$}
 \label{fig:fnlmassfcn}
\end{figure}

\begin{figure}[t]
\centering
\begin{tabular}{cc}
\includegraphics[width=80mm]{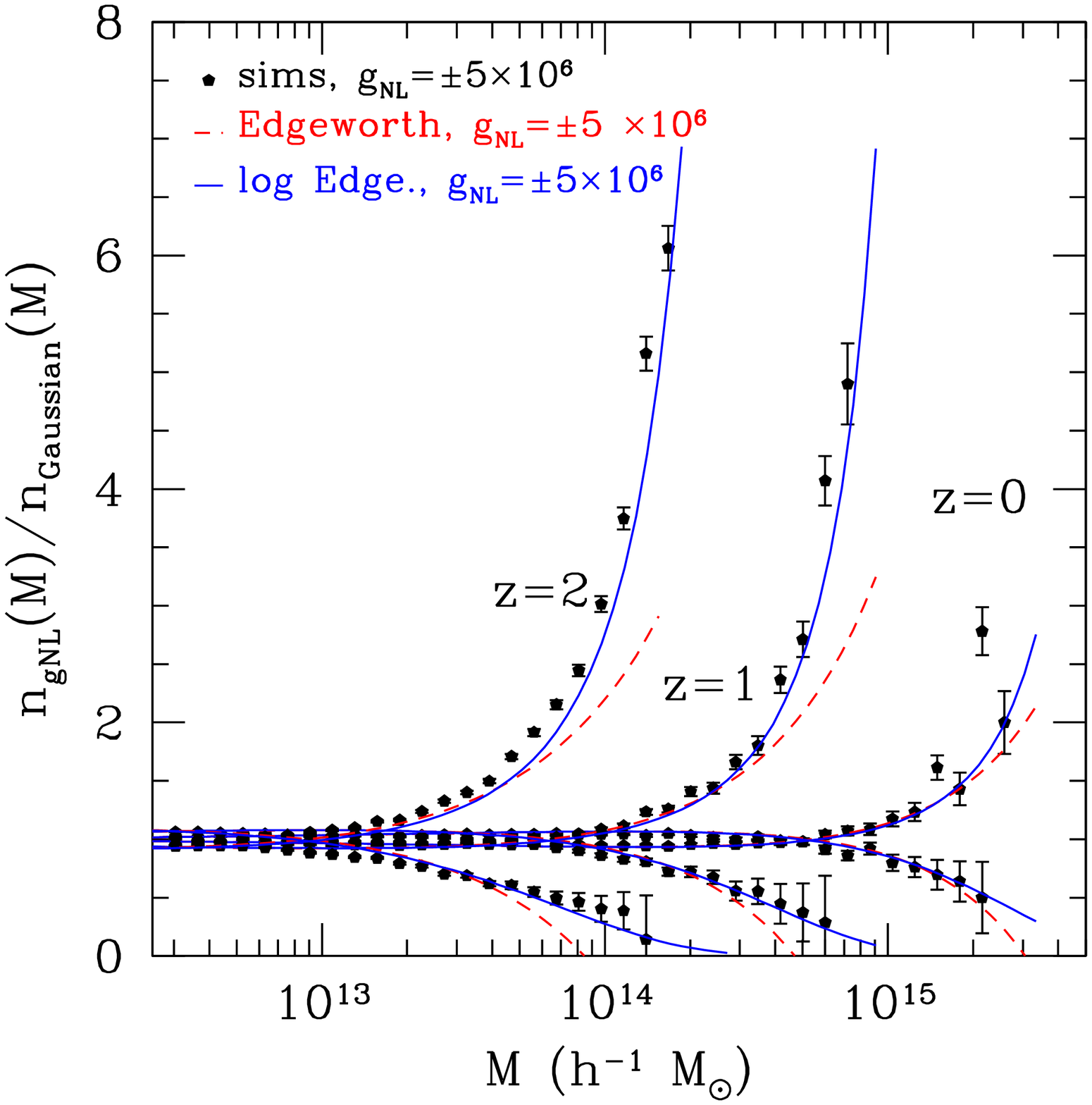} &\includegraphics[width=80mm]{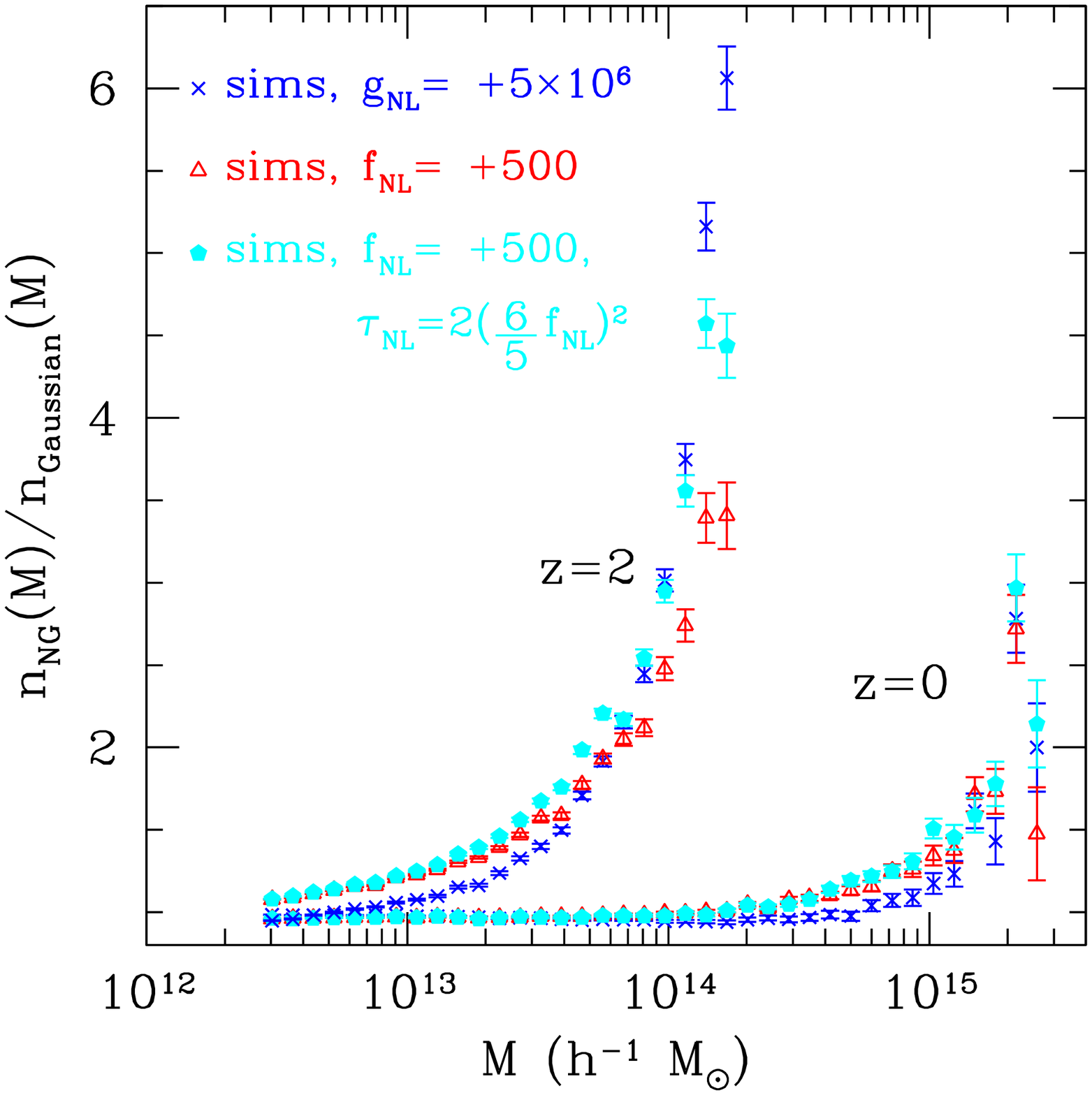}\\
\mbox{\bf(a)} & \mbox{\bf (b)}
\end{tabular}
\caption{The non-Gaussian correction to the halo mass function. Plotted is (a) $\fnl=0$, $\gnl=\pm5\times 10^{6}$.
 In panel (b) the non-Gaussian corrections for $(\fnl, \tau_{NL})=(500, (\frac{6}{5}500)^2)$, $(\fnl, \tau_{NL})=(500, 2(\frac{6}{5}500)^2)$ 
 and $\gnl=5\times 10^6$ are plotted together for comparison.} 
 \label{fig:gnlmassfcn}
\end{figure}

In Fig.~(\ref{fig:fnlmassfcn}) and Fig.~(\ref{fig:gnlmassfcn}) the ratio of the non-Gaussian mass function $n_{NG}(M)$ to Gaussian mass function $n_G(M)$ is plotted. The effects 
of primordial non-Gaussianity are clearly visible, positive (negative) $\fnl$, $\gnl$ increase (decrease) the abundance of halos, most significantly at high mass and high redshift.  The curves show the analytic predictions for the Edgeworth and log-Edgeworth mass functions described in \S\ref{ssec:massfcn}. In regions where the effects of 
non-Gaussianity are small, both are in reasonable agreement with the N-body results. However, at the high mass end where non-Gaussian effects
are largest the log-Edgeworth mass function is clearly in better agreement.  

In panel (b) of Fig.~(\ref{fig:gnlmassfcn}) the non-Gaussian corrections for ($\fnl$, $\tau_{NL}=(\frac{6}{5}\fnl)^2$), ($\fnl=0$, $\gnl$) and ($\fnl$, $\tau_{NL}=2(\frac{6}{5}\fnl)^2$),
are plotted together. For $\tau_{NL} \neq (\frac{6}{5}\fnl)^2$, the non-Gaussian effects are clearly larger than in the $\fnl$-only case. However, the mass and redshift dependence
of the curves is similar: we found that models with $\tau_{NL}\ge (\frac{6}{5}\fnl)^2$ can be made to look like models with $\tau_{NL}=(\frac{6}{5}\fnl)^2$ by using a larger value of $\fnl$. 
On the other hand, the mass dependence of $\gnl$ effects is distinctly steeper at high masses, thus in principle primordial skewness and kurtosis may be distinguishable with the mass function.

\begin{figure}
\centering
\begin{tabular}{cc}
\includegraphics[width=80mm]{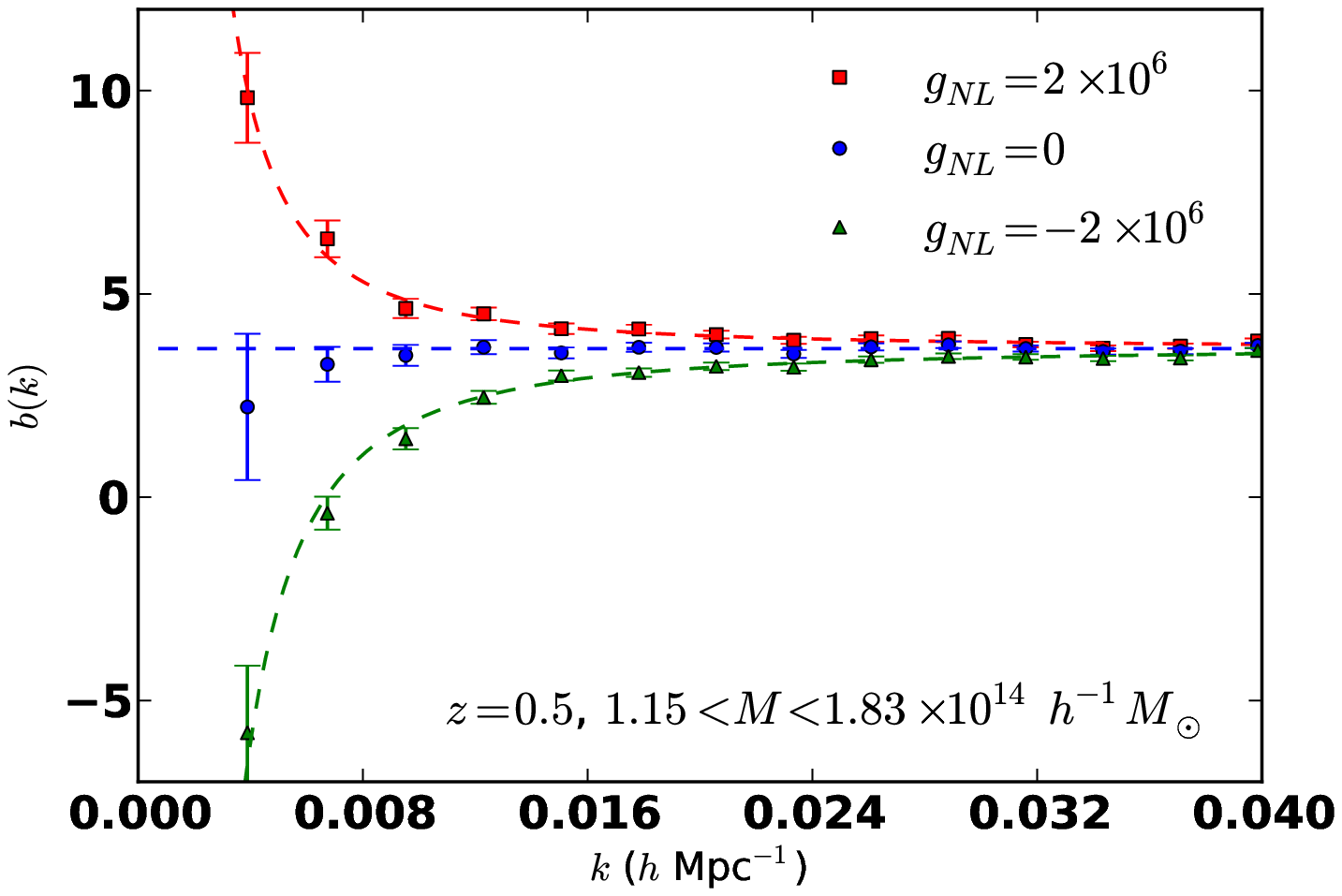}&\includegraphics[width=80mm]{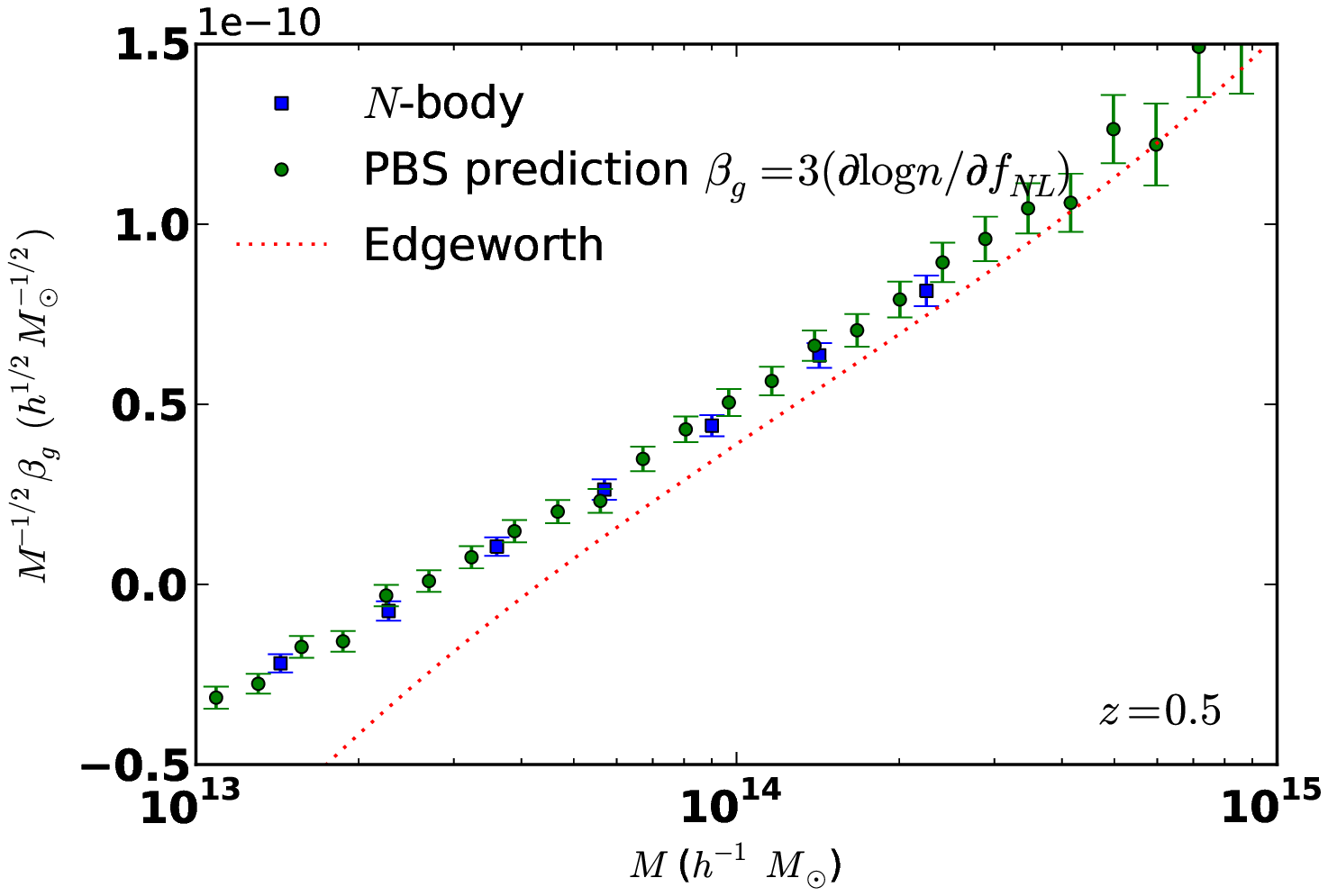}\\
\mbox{\bf(a)} & \mbox{\bf (b)}
\end{tabular}
\caption{(a) An example to illustrate that halo bias in a $\gnl$ cosmology takes the functional form
form $b(k) = b_1 + \beta_g \gnl/\alpha(k)$.  This figure corresponds to redshift $z=0.5$ and halo
mass range $1.15 \le M \le 1.83 \times 10^{14}$ $h^{-1}$ $M_\odot$, but we find the same functional
form for all redshifts and halo masses. (b) The mass dependence of the $\gnl$ bias, compared with the peak-background 
split prediction $3\frac{\partial \ln n}{\partial \fnl}$, there is excellent agreement.}
\label{fig:gnl}
\end{figure}

In Fig.~(\ref{fig:gnl}) we show the scale-dependent bias from $\gnl$ initial conditions. The $k$-dependence of the bias is accurately described by 
the form $b+\gnl\beta_g/\alpha(k)$ where $\beta_g$ is a constant. In panel (b) we compare the peak-background split prediction
$\beta_g=3\gnl\partial(\log n)/\partial \fnl$ with the value of $\beta_g$ measured from simulations. The agreement is excellent. Also shown 
is the analytic prediction for $3\partial(\log n)/\partial \fnl$ using the Edgeworth mass function. At these masses the analytic form is not 
sufficiently accurate to describe the $\gnl$ bias. 

\begin{figure}
\centering
\begin{tabular}{cc}
\includegraphics[width=80mm]{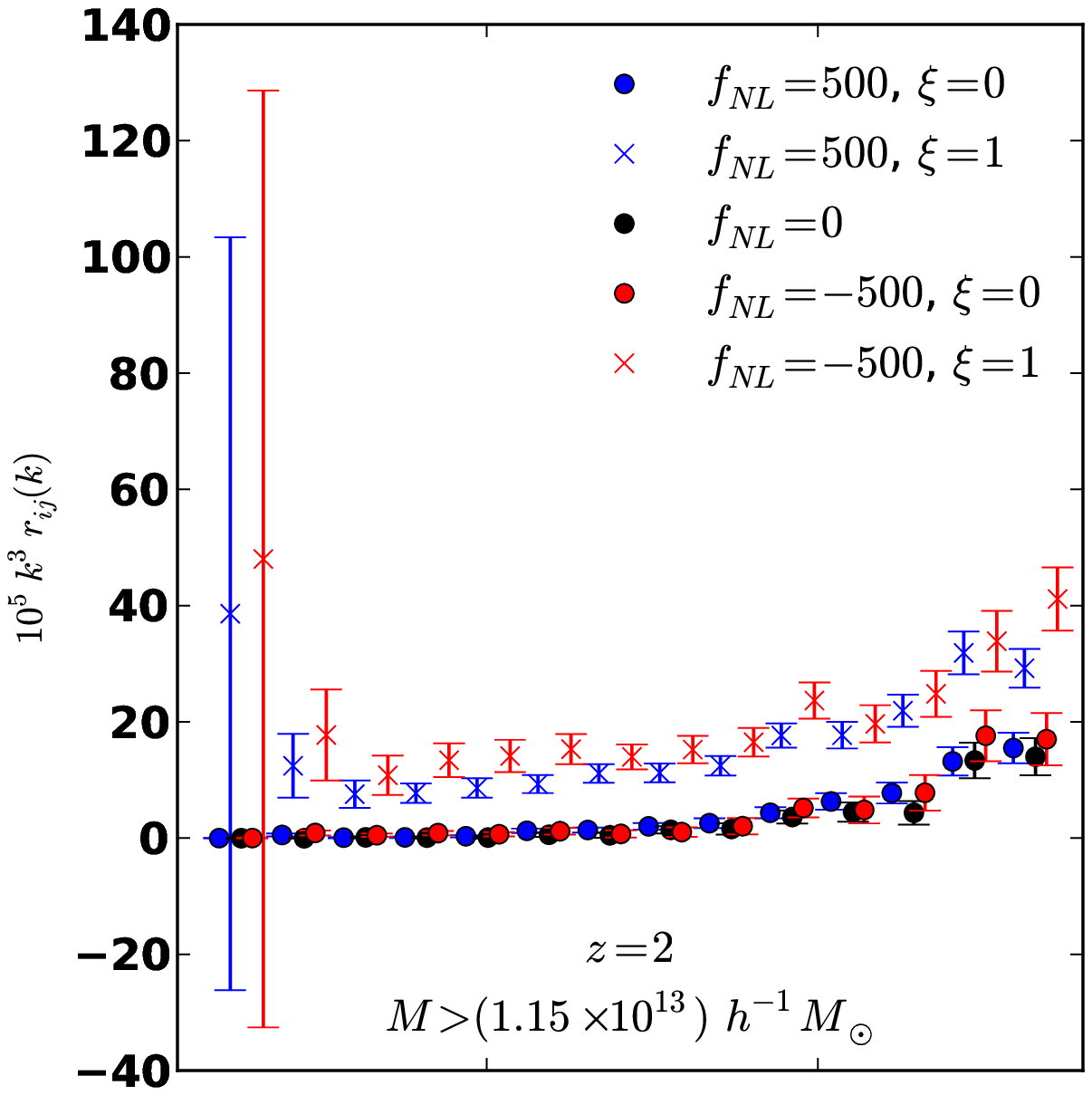}&\includegraphics[width=80mm]{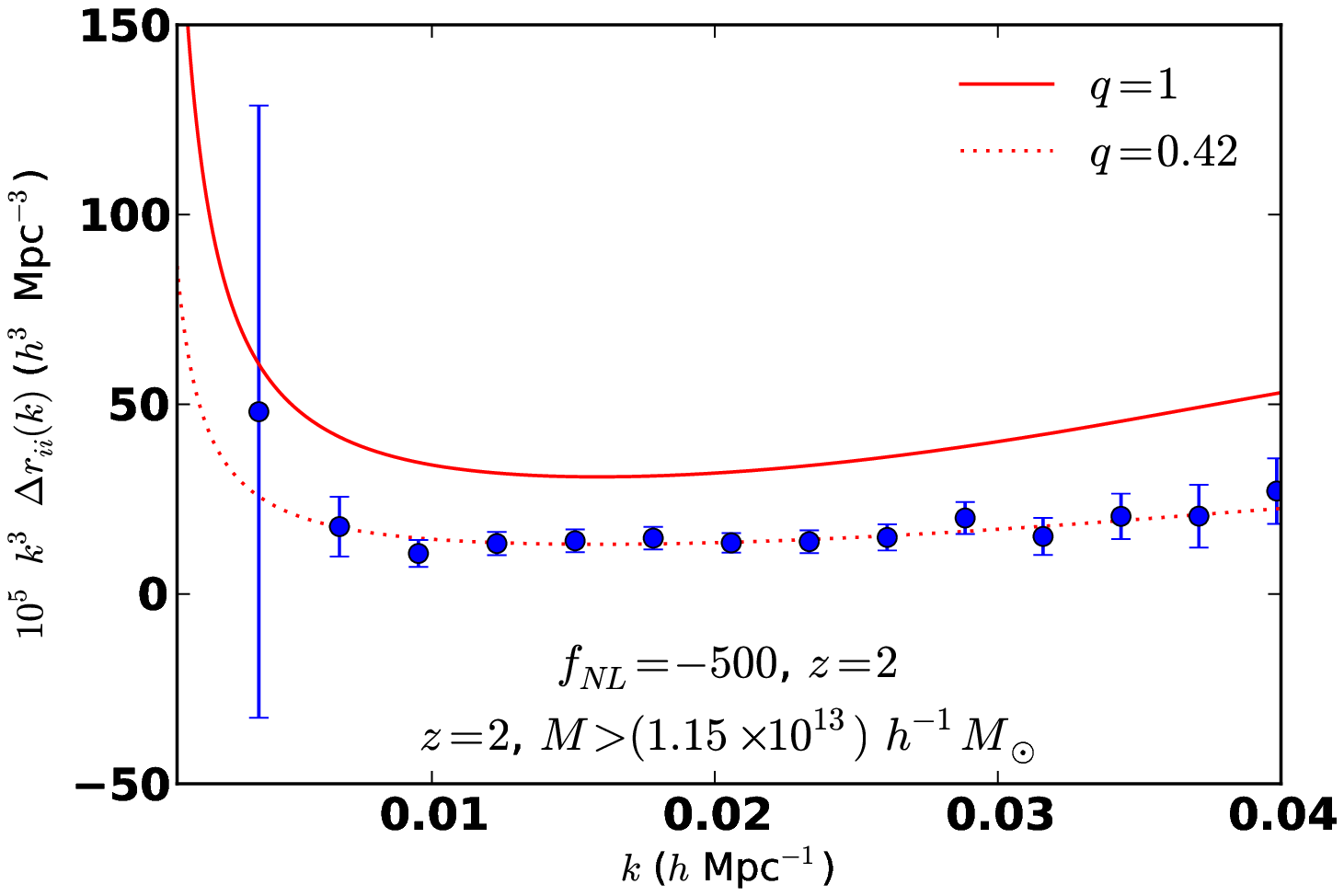}\\
\mbox{\bf(a)} & \mbox{\bf (b)}
\end{tabular}
\caption{(a) An example to illustrate that halo bias in the two-field model of \S \protect{\ref{ssec:tnl}} with $\tau_{NL}=(1+\xi^2)(\frac{6}{5}\fnl)^2$ is stochastic  ($r_{ij} \neq 0$). 
This is in contrast to the Gaussian case and the  $\fnl$-only model given in \S \ref{ssec:fnl}, for which halo bias is non-stochastic on large-scales ($r_{ij}=0$). 
This figure corresponds to redshift $z=0.5$ and halo
mass range $1.15 \le M \le 1.83 \times 10^{14}$ $h^{-1}$ $M_\odot$, but we find the same functional
form for all redshifts and halo masses. (b) An example of the peak-background split prediction (solid curve) with the measured
stochasticity -- in this case the lowest-order peak-background split clearly over-predicts the stochasticity.}
\label{fig:stochastic}
\end{figure}

We now consider the stochasticity induced by initial conditions with $\tau_{NL}\neq (\frac{6}{5}\fnl)^2$. The stochasticity parameter is given by,
\be
r_{ij}(k)=\frac{P_{h_ih_j}(k)-\delta_{ij}/n}{P_{mm}(k)}-\frac{P_{mh_i}(k)P_{mh_j}}{P_{mm}^2(k)}\,.
\ee
where $h_i$ stands for halos in the $i^{th}$ mass bin. 
Stochasticity is generally expected on small scales from the $1$-halo term in the halo power spectrum. But on large scales
halos are expected to be non-stochastic tracers of the dark matter. In Fig.~(\ref{fig:stochastic})
we show that the stochasticity vanishes on large-scales for Gaussian initial conditions and $\fnl$ initial conditions described by Eq.~(\ref{eq:fnldef}). 
On the other hand, the two field initial conditions described in Eq.~(\ref{eq:tnldef}) giving $\tau_{NL}\neq (\frac{6}{5}\fnl)^2$ do give rise to large-scale 
stochasticity. 

In panel (b), we show the scale dependence of the stochasticity in comparison with the lowest-order peak-background split prediction
 from \S \ref{ssec:stochastic}. Here we find disagreement: the scale dependence of the analytic prediction agrees well with what is seen in simulations
 but the amplitude is too high by $\sim 50\%$. The disagreement between the amplitude of the stochasticity seen in simulations and the prediction from Eq.~(\ref{eq:Phhtnl})
varies with mass and redshift, but is typically $\sim 30\%$ \cite{Smith:2010gx}. For Gaussian initial conditions, we also found inconsistent agreement between the 
halo model predictions for $r_{ii}$ and the measured values. 
This mismatch in $r_{ii}$ values is in qualitative agreement with \cite{Hamaus:2010im}.

\section{Summary}
\label{sec:summary}

We have considered the impact of three types of primordial non-Gaussianity, described by parameters $\fnl$, $\gnl$ and $\tau_{NL}$, 
on the abundance and clustering of dark matter halos. Analytic predictions for the halo abundance that are based on using the Edgeworth series 
for the non-Gaussian PDF in the Press-Schechter model agree well with simulations, provided the ``log-Edgeworth" truncation is used (see Figs.~(\ref{fig:fnlmassfcn}), (\ref{fig:gnlmassfcn}) \& \cite{LoVerde:2011iz}).
We found a simple peak-background split description of halo bias from $\gnl$ initial conditions (Eq.~(\ref{eq:bkfg})) that is in excellent agreement with simulations 
(see Fig.~(\ref{fig:gnl}) \& \cite{Smith:2011ub}).  Our simulations show
that the two-field model given in \S\ref{ssec:tnl} that gives rise to $\tau_{NL}\neq (\frac{6}{5}\fnl)^2$, does indeed generate large scale stochasticity. 
Unfortunately, the analytic description in \S\ref{ssec:stochastic} predicts only the $k$-dependence of the stochasticity accurately, 
the amplitude is incorrect at the $\lsim 50\%$ level (see Fig.~(\ref{fig:stochastic}) \& \cite{Smith:2010gx}). We conclude that scale-dependent bias from 
$\fnl$ and $\gnl$ initial conditions is well-understood analytically. 
On the other hand, modeling halo stochasticity appears to be more difficult and the analytic prescription given in \S \ref{ssec:stochastic} is insufficiently accurate to 
interpret data without input from simulations \cite{Smith:2010gx}. 

\begin{acknowledgments}
 ML gratefully acknowledges support from the Institute for Advanced Study, the 
Friends of the Institute for Advanced Study and the NSF through AST-0807444. 
SF is supported through the Martin Schwarzschild fund in Astronomy at Princeton University.
KMS is supported by a Lyman Spitzer fellowship in the Department of Astrophysical Sciences at Princeton University.
This document is adapted from the instructions provided to the authors
of the proceedings papers at CHARM~07, Ithaca, NY~\cite{charm07},  
and from eConf templates~\cite{templates-ref}.
\end{acknowledgments}

\bigskip 
\bibliography{ngac}

\begin{thebibliography}{41}
\expandafter\ifx\csname natexlab\endcsname\relax\def\natexlab#1{#1}\fi
\expandafter\ifx\csname bibnamefont\endcsname\relax
  \def\bibnamefont#1{#1}\fi
\expandafter\ifx\csname bibfnamefont\endcsname\relax
  \def\bibfnamefont#1{#1}\fi
\expandafter\ifx\csname citenamefont\endcsname\relax
  \def\citenamefont#1{#1}\fi
\expandafter\ifx\csname url\endcsname\relax
  \def\url#1{\texttt{#1}}\fi
\expandafter\ifx\csname urlprefix\endcsname\relax\def\urlprefix{URL }\fi
\providecommand{\bibinfo}[2]{#2}
\providecommand{\eprint}[2][]{\url{#2}}

\bibitem[{\citenamefont{Guth and Pi}(1982)}]{Guth:1982ec}
\bibinfo{author}{\bibfnamefont{A.~H.} \bibnamefont{Guth}} \bibnamefont{and}
  \bibinfo{author}{\bibfnamefont{S.~Y.} \bibnamefont{Pi}},
  \bibinfo{journal}{Phys. Rev. Lett.} \textbf{\bibinfo{volume}{49}},
  \bibinfo{pages}{1110} (\bibinfo{year}{1982}).

\bibitem[{\citenamefont{Hawking}(1982)}]{Hawking:1982cz}
\bibinfo{author}{\bibfnamefont{S.~W.} \bibnamefont{Hawking}},
  \bibinfo{journal}{Phys. Lett.} \textbf{\bibinfo{volume}{B115}},
  \bibinfo{pages}{295} (\bibinfo{year}{1982}).

\bibitem[{\citenamefont{Starobinsky}(1982)}]{Starobinsky:1982ee}
\bibinfo{author}{\bibfnamefont{A.~A.} \bibnamefont{Starobinsky}},
  \bibinfo{journal}{Phys. Lett.} \textbf{\bibinfo{volume}{B117}},
  \bibinfo{pages}{175} (\bibinfo{year}{1982}).

\bibitem[{\citenamefont{Bardeen et~al.}(1983)\citenamefont{Bardeen, Steinhardt,
  and Turner}}]{Bardeen:1983qw}
\bibinfo{author}{\bibfnamefont{J.~M.} \bibnamefont{Bardeen}},
  \bibinfo{author}{\bibfnamefont{P.~J.} \bibnamefont{Steinhardt}},
  \bibnamefont{and} \bibinfo{author}{\bibfnamefont{M.~S.}
  \bibnamefont{Turner}}, \bibinfo{journal}{Phys. Rev.}
  \textbf{\bibinfo{volume}{D28}}, \bibinfo{pages}{679} (\bibinfo{year}{1983}).

\bibitem[{\citenamefont{Komatsu et~al.}(2011)}]{Komatsu:2010fb}
\bibinfo{author}{\bibfnamefont{E.}~\bibnamefont{Komatsu}} \bibnamefont{et~al.},
  \bibinfo{journal}{Astrophys. J. Suppl.} \textbf{\bibinfo{volume}{192}},
  \bibinfo{pages}{18} (\bibinfo{year}{2011}), \eprint{1001.4538}.

\bibitem[{\citenamefont{Bartolo et~al.}(2004)\citenamefont{Bartolo, Komatsu,
  Matarrese, and Riotto}}]{Bartolo:2004if}
\bibinfo{author}{\bibfnamefont{N.}~\bibnamefont{Bartolo}},
  \bibinfo{author}{\bibfnamefont{E.}~\bibnamefont{Komatsu}},
  \bibinfo{author}{\bibfnamefont{S.}~\bibnamefont{Matarrese}},
  \bibnamefont{and} \bibinfo{author}{\bibfnamefont{A.}~\bibnamefont{Riotto}},
  \bibinfo{journal}{Phys.Rept.} \textbf{\bibinfo{volume}{402}},
  \bibinfo{pages}{103} (\bibinfo{year}{2004}), \eprint{astro-ph/0406398}.

\bibitem[{\citenamefont{Chen}(2010)}]{Chen:2010xka}
\bibinfo{author}{\bibfnamefont{X.}~\bibnamefont{Chen}}, \bibinfo{journal}{Adv.
  Astron.} \textbf{\bibinfo{volume}{2010}}, \bibinfo{pages}{638979}
  (\bibinfo{year}{2010}), \eprint{1002.1416}.

\bibitem[{\citenamefont{Fergusson et~al.}(2010)\citenamefont{Fergusson, Regan,
  and Shellard}}]{Fergusson:2010gn}
\bibinfo{author}{\bibfnamefont{J.~R.} \bibnamefont{Fergusson}},
  \bibinfo{author}{\bibfnamefont{D.~M.} \bibnamefont{Regan}}, \bibnamefont{and}
  \bibinfo{author}{\bibfnamefont{E.~P.~S.} \bibnamefont{Shellard}}
  (\bibinfo{year}{2010}), \eprint{1012.6039}.

\bibitem[{\citenamefont{Smidt et~al.}(2010)}]{Smidt:2010sv}
\bibinfo{author}{\bibfnamefont{J.}~\bibnamefont{Smidt}} \bibnamefont{et~al.}
  (\bibinfo{year}{2010}), \eprint{1001.5026}.

\bibitem[{\citenamefont{Slosar et~al.}(2008)\citenamefont{Slosar, Hirata,
  Seljak, Ho, and Padmanabhan}}]{Slosar:2008hx}
\bibinfo{author}{\bibfnamefont{A.}~\bibnamefont{Slosar}},
  \bibinfo{author}{\bibfnamefont{C.}~\bibnamefont{Hirata}},
  \bibinfo{author}{\bibfnamefont{U.}~\bibnamefont{Seljak}},
  \bibinfo{author}{\bibfnamefont{S.}~\bibnamefont{Ho}}, \bibnamefont{and}
  \bibinfo{author}{\bibfnamefont{N.}~\bibnamefont{Padmanabhan}},
  \bibinfo{journal}{JCAP} \textbf{\bibinfo{volume}{0808}}, \bibinfo{pages}{031}
  (\bibinfo{year}{2008}), \eprint{0805.3580}.

\bibitem[{\citenamefont{Carbone et~al.}(2010)\citenamefont{Carbone, Mena, and
  Verde}}]{Carbone:2010sb}
\bibinfo{author}{\bibfnamefont{C.}~\bibnamefont{Carbone}},
  \bibinfo{author}{\bibfnamefont{O.}~\bibnamefont{Mena}}, \bibnamefont{and}
  \bibinfo{author}{\bibfnamefont{L.}~\bibnamefont{Verde}},
  \bibinfo{journal}{JCAP} \textbf{\bibinfo{volume}{1007}}, \bibinfo{pages}{020}
  (\bibinfo{year}{2010}), \eprint{1003.0456}.

\bibitem[{\citenamefont{Shandera et~al.}(2010)\citenamefont{Shandera, Dalal,
  and Huterer}}]{Shandera:2010ei}
\bibinfo{author}{\bibfnamefont{S.}~\bibnamefont{Shandera}},
  \bibinfo{author}{\bibfnamefont{N.}~\bibnamefont{Dalal}}, \bibnamefont{and}
  \bibinfo{author}{\bibfnamefont{D.}~\bibnamefont{Huterer}}
  (\bibinfo{year}{2010}), \eprint{1010.3722}.

\bibitem[{\citenamefont{LoVerde and Smith}(2011)}]{LoVerde:2011iz}
\bibinfo{author}{\bibfnamefont{M.}~\bibnamefont{LoVerde}} \bibnamefont{and}
  \bibinfo{author}{\bibfnamefont{K.~M.} \bibnamefont{Smith}},
  \bibinfo{journal}{JCAP} \textbf{\bibinfo{volume}{1108}}, \bibinfo{pages}{003}
  (\bibinfo{year}{2011}), \eprint{1102.1439}.

\bibitem[{\citenamefont{Smith and LoVerde}(2010)}]{Smith:2010gx}
\bibinfo{author}{\bibfnamefont{K.~M.} \bibnamefont{Smith}} \bibnamefont{and}
  \bibinfo{author}{\bibfnamefont{M.}~\bibnamefont{LoVerde}}
  (\bibinfo{year}{2010}), \eprint{1010.0055}.

\bibitem[{\citenamefont{Smith et~al.}(2011{\natexlab{a}})\citenamefont{Smith,
  Ferraro, and LoVerde}}]{Smith:2011ub}
\bibinfo{author}{\bibfnamefont{K.~M.} \bibnamefont{Smith}},
  \bibinfo{author}{\bibfnamefont{S.}~\bibnamefont{Ferraro}}, \bibnamefont{and}
  \bibinfo{author}{\bibfnamefont{M.}~\bibnamefont{LoVerde}}
  (\bibinfo{year}{2011}{\natexlab{a}}), \eprint{1106.0503}.

\bibitem[{\citenamefont{Smith et~al.}(2011{\natexlab{b}})\citenamefont{Smith,
  LoVerde, and Zaldarriaga}}]{Smith:2011if}
\bibinfo{author}{\bibfnamefont{K.~M.} \bibnamefont{Smith}},
  \bibinfo{author}{\bibfnamefont{M.}~\bibnamefont{LoVerde}}, \bibnamefont{and}
  \bibinfo{author}{\bibfnamefont{M.}~\bibnamefont{Zaldarriaga}}
  (\bibinfo{year}{2011}{\natexlab{b}}), \eprint{1108.1805}.

\bibitem[{\citenamefont{Babich et~al.}(2004)\citenamefont{Babich, Creminelli,
  and Zaldarriaga}}]{Babich:2004gb}
\bibinfo{author}{\bibfnamefont{D.}~\bibnamefont{Babich}},
  \bibinfo{author}{\bibfnamefont{P.}~\bibnamefont{Creminelli}},
  \bibnamefont{and}
  \bibinfo{author}{\bibfnamefont{M.}~\bibnamefont{Zaldarriaga}},
  \bibinfo{journal}{JCAP} \textbf{\bibinfo{volume}{0408}}, \bibinfo{pages}{009}
  (\bibinfo{year}{2004}), \eprint{astro-ph/0405356}.

\bibitem[{\citenamefont{Linde and Mukhanov}(1997)}]{Linde:1996gt}
\bibinfo{author}{\bibfnamefont{A.~D.} \bibnamefont{Linde}} \bibnamefont{and}
  \bibinfo{author}{\bibfnamefont{V.~F.} \bibnamefont{Mukhanov}},
  \bibinfo{journal}{Phys. Rev.} \textbf{\bibinfo{volume}{D56}},
  \bibinfo{pages}{535} (\bibinfo{year}{1997}), \eprint{astro-ph/9610219}.

\bibitem[{\citenamefont{Lyth and Wands}(2002)}]{Lyth:2001nq}
\bibinfo{author}{\bibfnamefont{D.~H.} \bibnamefont{Lyth}} \bibnamefont{and}
  \bibinfo{author}{\bibfnamefont{D.}~\bibnamefont{Wands}},
  \bibinfo{journal}{Phys. Lett.} \textbf{\bibinfo{volume}{B524}},
  \bibinfo{pages}{5} (\bibinfo{year}{2002}), \eprint{hep-ph/0110002}.

\bibitem[{\citenamefont{Byrnes and Choi}(2010)}]{Byrnes:2010em}
\bibinfo{author}{\bibfnamefont{C.~T.} \bibnamefont{Byrnes}} \bibnamefont{and}
  \bibinfo{author}{\bibfnamefont{K.-Y.} \bibnamefont{Choi}},
  \bibinfo{journal}{Adv.Astron.} \textbf{\bibinfo{volume}{2010}},
  \bibinfo{pages}{724525} (\bibinfo{year}{2010}), \bibinfo{note}{* Temporary
  entry *}, \eprint{1002.3110}.

\bibitem[{\citenamefont{Komatsu and Spergel}(2001)}]{Komatsu:2001rj}
\bibinfo{author}{\bibfnamefont{E.}~\bibnamefont{Komatsu}} \bibnamefont{and}
  \bibinfo{author}{\bibfnamefont{D.~N.} \bibnamefont{Spergel}},
  \bibinfo{journal}{Phys. Rev.} \textbf{\bibinfo{volume}{D63}},
  \bibinfo{pages}{063002} (\bibinfo{year}{2001}), \eprint{astro-ph/0005036}.

\bibitem[{\citenamefont{Okamoto and Hu}(2002)}]{Okamoto:2002ik}
\bibinfo{author}{\bibfnamefont{T.}~\bibnamefont{Okamoto}} \bibnamefont{and}
  \bibinfo{author}{\bibfnamefont{W.}~\bibnamefont{Hu}}, \bibinfo{journal}{Phys.
  Rev.} \textbf{\bibinfo{volume}{D66}}, \bibinfo{pages}{063008}
  (\bibinfo{year}{2002}), \eprint{astro-ph/0206155}.

\bibitem[{\citenamefont{Enqvist and Nurmi}(2005)}]{Enqvist:2005pg}
\bibinfo{author}{\bibfnamefont{K.}~\bibnamefont{Enqvist}} \bibnamefont{and}
  \bibinfo{author}{\bibfnamefont{S.}~\bibnamefont{Nurmi}},
  \bibinfo{journal}{JCAP} \textbf{\bibinfo{volume}{0510}}, \bibinfo{pages}{013}
  (\bibinfo{year}{2005}), \eprint{astro-ph/0508573}.

\bibitem[{\citenamefont{Boubekeur and Lyth}(2006)}]{Boubekeur:2005fj}
\bibinfo{author}{\bibfnamefont{L.}~\bibnamefont{Boubekeur}} \bibnamefont{and}
  \bibinfo{author}{\bibfnamefont{D.~H.} \bibnamefont{Lyth}},
  \bibinfo{journal}{Phys. Rev.} \textbf{\bibinfo{volume}{D73}},
  \bibinfo{pages}{021301} (\bibinfo{year}{2006}), \eprint{astro-ph/0504046}.

\bibitem[{\citenamefont{Suyama and Yamaguchi}(2008)}]{Suyama:2007bg}
\bibinfo{author}{\bibfnamefont{T.}~\bibnamefont{Suyama}} \bibnamefont{and}
  \bibinfo{author}{\bibfnamefont{M.}~\bibnamefont{Yamaguchi}},
  \bibinfo{journal}{Phys. Rev.} \textbf{\bibinfo{volume}{D77}},
  \bibinfo{pages}{023505} (\bibinfo{year}{2008}), \eprint{0709.2545}.

\bibitem[{\citenamefont{Dunkley et~al.}(2009)}]{Dunkley:2008ie}
\bibinfo{author}{\bibfnamefont{J.}~\bibnamefont{Dunkley}} \bibnamefont{et~al.}
  (\bibinfo{collaboration}{WMAP}), \bibinfo{journal}{Astrophys. J. Suppl.}
  \textbf{\bibinfo{volume}{180}}, \bibinfo{pages}{306} (\bibinfo{year}{2009}),
  \eprint{0803.0586}.

\bibitem[{\citenamefont{Press and Schechter}(1974)}]{Press:1973iz}
\bibinfo{author}{\bibfnamefont{W.~H.} \bibnamefont{Press}} \bibnamefont{and}
  \bibinfo{author}{\bibfnamefont{P.}~\bibnamefont{Schechter}},
  \bibinfo{journal}{Astrophys. J.} \textbf{\bibinfo{volume}{187}},
  \bibinfo{pages}{425} (\bibinfo{year}{1974}).

\bibitem[{\citenamefont{Jenkins et~al.}(2001)}]{Jenkins:2000bv}
\bibinfo{author}{\bibfnamefont{A.}~\bibnamefont{Jenkins}} \bibnamefont{et~al.},
  \bibinfo{journal}{Mon. Not. Roy. Astron. Soc.}
  \textbf{\bibinfo{volume}{321}}, \bibinfo{pages}{372} (\bibinfo{year}{2001}),
  \eprint{astro-ph/0005260}.

\bibitem[{\citenamefont{Lucchin and Matarrese}(1988)}]{Lucchin:1987yv}
\bibinfo{author}{\bibfnamefont{F.}~\bibnamefont{Lucchin}} \bibnamefont{and}
  \bibinfo{author}{\bibfnamefont{S.}~\bibnamefont{Matarrese}},
  \bibinfo{journal}{Astrophys. J.} \textbf{\bibinfo{volume}{330}},
  \bibinfo{pages}{535} (\bibinfo{year}{1988}).

\bibitem[{\citenamefont{Matarrese et~al.}(2000)\citenamefont{Matarrese, Verde,
  and Jimenez}}]{Matarrese:2000iz}
\bibinfo{author}{\bibfnamefont{S.}~\bibnamefont{Matarrese}},
  \bibinfo{author}{\bibfnamefont{L.}~\bibnamefont{Verde}}, \bibnamefont{and}
  \bibinfo{author}{\bibfnamefont{R.}~\bibnamefont{Jimenez}},
  \bibinfo{journal}{Astrophys. J.} \textbf{\bibinfo{volume}{541}},
  \bibinfo{pages}{10} (\bibinfo{year}{2000}), \eprint{astro-ph/0001366}.

\bibitem[{\citenamefont{LoVerde et~al.}(2008)\citenamefont{LoVerde, Miller,
  Shandera, and Verde}}]{LoVerde:2007ri}
\bibinfo{author}{\bibfnamefont{M.}~\bibnamefont{LoVerde}},
  \bibinfo{author}{\bibfnamefont{A.}~\bibnamefont{Miller}},
  \bibinfo{author}{\bibfnamefont{S.}~\bibnamefont{Shandera}}, \bibnamefont{and}
  \bibinfo{author}{\bibfnamefont{L.}~\bibnamefont{Verde}},
  \bibinfo{journal}{JCAP} \textbf{\bibinfo{volume}{0804}}, \bibinfo{pages}{014}
  (\bibinfo{year}{2008}), \eprint{0711.4126}.

\bibitem[{\citenamefont{Pillepich et~al.}(2008)\citenamefont{Pillepich,
  Porciani, and Hahn}}]{Pillepich:2008ka}
\bibinfo{author}{\bibfnamefont{A.}~\bibnamefont{Pillepich}},
  \bibinfo{author}{\bibfnamefont{C.}~\bibnamefont{Porciani}}, \bibnamefont{and}
  \bibinfo{author}{\bibfnamefont{O.}~\bibnamefont{Hahn}}
  (\bibinfo{year}{2008}), \eprint{0811.4176}.

\bibitem[{\citenamefont{Dalal et~al.}(2008)\citenamefont{Dalal, Dore, Huterer,
  and Shirokov}}]{Dalal:2007cu}
\bibinfo{author}{\bibfnamefont{N.}~\bibnamefont{Dalal}},
  \bibinfo{author}{\bibfnamefont{O.}~\bibnamefont{Dore}},
  \bibinfo{author}{\bibfnamefont{D.}~\bibnamefont{Huterer}}, \bibnamefont{and}
  \bibinfo{author}{\bibfnamefont{A.}~\bibnamefont{Shirokov}},
  \bibinfo{journal}{Phys. Rev.} \textbf{\bibinfo{volume}{D77}},
  \bibinfo{pages}{123514} (\bibinfo{year}{2008}), \eprint{0710.4560}.

\bibitem[{\citenamefont{Desjacques and Seljak}(2010)}]{Desjacques:2009jb}
\bibinfo{author}{\bibfnamefont{V.}~\bibnamefont{Desjacques}} \bibnamefont{and}
  \bibinfo{author}{\bibfnamefont{U.}~\bibnamefont{Seljak}},
  \bibinfo{journal}{Phys. Rev.} \textbf{\bibinfo{volume}{D81}},
  \bibinfo{pages}{023006} (\bibinfo{year}{2010}), \eprint{0907.2257}.

\bibitem[{\citenamefont{Desjacques et~al.}(2011)\citenamefont{Desjacques,
  Jeong, and Schmidt}}]{Desjacques:2011mq}
\bibinfo{author}{\bibfnamefont{V.}~\bibnamefont{Desjacques}},
  \bibinfo{author}{\bibfnamefont{D.}~\bibnamefont{Jeong}}, \bibnamefont{and}
  \bibinfo{author}{\bibfnamefont{F.}~\bibnamefont{Schmidt}}
  (\bibinfo{year}{2011}), \eprint{1105.3628}.

\bibitem[{\citenamefont{Grossi et~al.}(2009)}]{Grossi:2009an}
\bibinfo{author}{\bibfnamefont{M.}~\bibnamefont{Grossi}} \bibnamefont{et~al.},
  \bibinfo{journal}{Mon. Not. Roy. Astron. Soc.}
  \textbf{\bibinfo{volume}{398}}, \bibinfo{pages}{321} (\bibinfo{year}{2009}),
  \eprint{0902.2013}.

\bibitem[{\citenamefont{Tseliakhovich et~al.}(2010)\citenamefont{Tseliakhovich,
  Hirata, and Slosar}}]{Tseliakhovich:2010kf}
\bibinfo{author}{\bibfnamefont{D.}~\bibnamefont{Tseliakhovich}},
  \bibinfo{author}{\bibfnamefont{C.}~\bibnamefont{Hirata}}, \bibnamefont{and}
  \bibinfo{author}{\bibfnamefont{A.}~\bibnamefont{Slosar}}
  (\bibinfo{year}{2010}), \eprint{1004.3302}.

\bibitem[{\citenamefont{Springel}(2005)}]{Springel:2005mi}
\bibinfo{author}{\bibfnamefont{V.}~\bibnamefont{Springel}},
  \bibinfo{journal}{Mon. Not. Roy. Astron. Soc.}
  \textbf{\bibinfo{volume}{364}}, \bibinfo{pages}{1105} (\bibinfo{year}{2005}),
  \eprint{astro-ph/0505010}.

\bibitem[{\citenamefont{Lewis et~al.}(2000)\citenamefont{Lewis, Challinor, and
  Lasenby}}]{Lewis:1999bs}
\bibinfo{author}{\bibfnamefont{A.}~\bibnamefont{Lewis}},
  \bibinfo{author}{\bibfnamefont{A.}~\bibnamefont{Challinor}},
  \bibnamefont{and} \bibinfo{author}{\bibfnamefont{A.}~\bibnamefont{Lasenby}},
  \bibinfo{journal}{Astrophys. J.} \textbf{\bibinfo{volume}{538}},
  \bibinfo{pages}{473} (\bibinfo{year}{2000}), \eprint{astro-ph/9911177}.

\bibitem[{\citenamefont{Frenk et~al.}(1988)\citenamefont{Frenk, White, Davis,
  and Efstathiou}}]{Frenk:1988zz}
\bibinfo{author}{\bibfnamefont{C.~S.} \bibnamefont{Frenk}},
  \bibinfo{author}{\bibfnamefont{S.~D.~M.} \bibnamefont{White}},
  \bibinfo{author}{\bibfnamefont{M.}~\bibnamefont{Davis}}, \bibnamefont{and}
  \bibinfo{author}{\bibfnamefont{G.}~\bibnamefont{Efstathiou}},
  \bibinfo{journal}{Astrophys. J.} \textbf{\bibinfo{volume}{327}},
  \bibinfo{pages}{507} (\bibinfo{year}{1988}).

\bibitem[{\citenamefont{Hamaus et~al.}(2010)\citenamefont{Hamaus, Seljak,
  Desjacques, Smith, and Baldauf}}]{Hamaus:2010im}
\bibinfo{author}{\bibfnamefont{N.}~\bibnamefont{Hamaus}},
  \bibinfo{author}{\bibfnamefont{U.}~\bibnamefont{Seljak}},
  \bibinfo{author}{\bibfnamefont{V.}~\bibnamefont{Desjacques}},
  \bibinfo{author}{\bibfnamefont{R.~E.} \bibnamefont{Smith}}, \bibnamefont{and}
  \bibinfo{author}{\bibfnamefont{T.}~\bibnamefont{Baldauf}}
  (\bibinfo{year}{2010}), \eprint{1004.5377}.
\bibitem{charm07}   http://www.lepp.cornell.edu/charm07/

\bibitem{templates-ref} http://www.slac.stanford.edu/econf/editors/eprint-template/instructions.html

\end{thebibliography}


\end{document}